\documentclass[twocolumn,showpacs,preprintnumbers,amsmath,amssymb]{revtex4}

\usepackage{graphicx}
\usepackage{dcolumn}
\usepackage{bm}

\begin{document}

\title{Exact enumeration of the Critical States in the Oslo Model}

\author{Alvin Chua}

\author{Kim Christensen$^*$}
\affiliation{Blackett Laboratory, Imperial College, Prince Consort Road, London SW7 2BW, United Kingdom}

\date{\today}

\begin{abstract}
We determine analytically the number $N_{\mathcal{R}}(L)$ of recurrent states
in the $1d$ Oslo model as a function of system size $L$.
The solution
$N_{\mathcal{R}}(L) = \frac{1+\sqrt{5}}{2\sqrt{5}}\left(\frac{3+\sqrt{5}}{2}\right)^L 
+ \frac{\sqrt{5}-1}{2\sqrt{5}}\left(\frac{3-\sqrt{5}}{2}\right)^L$ is in
exact agreement with the number enumerated in computer simulations for 
$L = 1 - 10$.
For $L \gg 1$, the number of allowed metastable states in the
attractor increases exponentially as
$N_{\mathcal{R}}(L) \approx c_+ {\lambda}_+^L$,
where $\lambda_+ = \frac{3+\sqrt{5}}{2}$ is the golden mean. 
The system is non-ergodic in the sense that the states in the attractor
are not equally probable.
\end{abstract}

\maketitle

Some non-equilibrium systems can be described by simple physical laws.  
Despite this being the case, we find complex and non-linear behavior in many such systems.
A slowly driven granular system illustrates this phenomenon
spectacularly. If grains are deposited onto
a finite surface, they will eventually build into a pile.
The pile does not collapse and flatten due to the
friction between the grains.   
We call the different profiles of piles produced by this process, metastable
 states.
The addition of grains and the dissipation events (avalanches)
form a delicate feedback loop. 
Avalanches initiated by dropping
grains onto the pile will ensure that the slope
does not become too steep, while the addition of grains prevents
the slope from becoming too shallow. 
Eventually, the system settles into an attractor where,
the average avalanche size
of the system diverges in response to small perturbations induced
by adding grains.
The distribution of avalanche sizes becomes scale 
invariant and a
small perturbation may initiate small as well as large avalanches.
As such, the system is intrinsically non-linear and displays
characteristics similar to equilibrium
systems poised at a critical point. 
This has been observed experimentally, for real granular
systems where friction dominates over inertial effects \cite{Nature}, as
well as numerically \cite{BTW87,Oslo1}. It is an example
of systems displaying self-organized criticality \cite{BTW87, PBak,HJensen}.

We study the attractor of a simple $1d$ granular model 
known as the Oslo model. It successfully models
slowly driven granular systems displaying self-organized criticality
\cite{Nature,Oslo1}. The model 
has been shown to be a member of a large universality 
class of $1d$ systems, which includes the de-pinning transition of 
an interface dragged through a random medium and the deterministic
Burridge-Knopoff train model for earthquakes \cite{Maya,BK,Train}.  
Thus, the Oslo model plays the role of the ``Ising model'' for
$1d$ self-organized critical systems and has been
widely studied \cite{Maya,Train,Priezzhev,ref1,ref2,ref3,ref4,ref5,ref6,ref7,ref8,ref9,ref10,ref11}.
However, in contrast to various deterministic models of granular systems
where exact results do exist \cite{Dhar1, Dhar2}, the Oslo model
has resisted analytic treatment \cite{Priezzhev}.  
Only a handful of exact results are known for
non-equilibrium systems in general \cite{Derrida} 
and self-organized critical systems
in particular \cite{Dhar1,Dhar2,Tang}.
Any analytic result for the Oslo model would be
very valuable.

We are able to determine the number $N_{\mathcal{R}}(L)$ of states
in the attractor of the Oslo Model.
We show that
$N_{\mathcal{R}}(L) = \frac{1+\sqrt{5}}{2\sqrt{5}}\left(\frac{3+\sqrt{5}}{2}\right)^L 
+ \frac{\sqrt{5}-1}{2\sqrt{5}}\left(\frac{3-\sqrt{5}}{2}\right)^L$.
For system sizes $L \gg 1$, the number increases exponentially
fast as $N_{\mathcal{R}}(L) \approx c_+ {\lambda}_+^L$,
where $\lambda_+$ is the golden mean. 
This is a clear signature of the complexity displayed by this simple model.
Moreover, the system is non-ergodic. The recurrent states are not visited 
with the same frequency.

\begin{figure}[h]
\includegraphics[scale = 0.9]{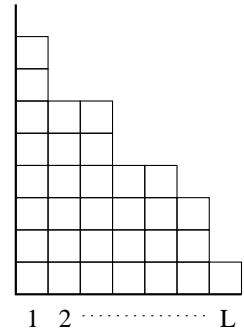}
\caption{The Oslo model of a $1d$ granular pile. An integer height
$h_i \geq 0$ is assigned to each site $i$. Grains are added at the
site $i=1$ next to the vertical wall by letting $h_1 \rightarrow h_1 + 1$.
Grains can leave the system at the open boundary, $h_L \rightarrow h_L - 1$. }
\label{Fig1}
\end{figure}

The Oslo model is defined on a one-dimensional discrete 
lattice consisting of $L$ sites,
$i = 1, 2, \ldots , L$. There is a vertical wall at the left boundary next to site $i=1$.
The system is open at the right boundary next to site $i=L$, where grains are allowed
to leave the system.
We will denote the respective heights of
the pile starting from the vertical wall by an $n$-tuple
$\mathbf{h}= \left(h_1,h_2,\ldots,h_i,\ldots,h_L\right)$,
see Fig. \ref{Fig1}.
In accordance with the experiment in Ref. \cite{Nature}, grains are
only inserted at $i = 1$, the site next to the vertical wall.
In the following discussion, it will be convenient to refer to local
slopes instead of heights. The slopes are denoted by
$\mathbf{z} =\left(z_1,z_2,\ldots,z_i,\ldots,z_L\right)$
with the definition
$z_i = h_i-h_{i+1}$ where $h_{L+1} = 0$.
Each site $i$ is assigned a critical slope,
${\bf z}^c(t) =\left(z^c_1(t),z^c_2(t), \ldots ,z^c_i(t), \ldots,z^c_L(t)\right)$.
If the local slope exceeds the critical slope,
$z_i > z_i^c(t)$, a grain at site $i$ will topple to the site $i+1$, that
is $h_i \rightarrow h_i -1, h_{i+1} \rightarrow h_{i+1} + 1$.
As a consequence, the slopes at neighboring sites
$i \pm 1$ may also exceed their critical slopes and topple, consequently  
causing an avalanche that propagates until the system settles in
a metastable state with $z_i \leq z_i^c$ for all $i$.

The critical slope $z_i^c(t)$ is a function of time as 
it is chosen randomly every time site $i$ topples.
We denote the minimum allowed value of the random 
critical slopes $z^{c(min)}$ and similarly the maximum allowed value
$z^{c(max)}$. 
In general, we can use any discrete probability distribution of 
critical slopes. However, we will consider the
simplest case where $z_i^c$ is chosen at random 
between the $1$ or $2$, $z_i^c(t) \in \{1,2\}$ when site $i$ topples,
such that $z^{c(min)}=1$ and $z^{c(max)}=2$.

The algorithm for the Oslo model is defined as follows.
\begin{enumerate}
\item 
Initialize the critical slopes $z_i^c$ and the
system in a arbitrary metastable state
with $z_i \leq z^c_i$ for all $i$.

\item 
Add a grain at site $i=1$, i.e., $z_1 \rightarrow z_1 + 1$.

\item 
If $z_i > z_i^c$, the site relaxes and
\begin{eqnarray*}
z_i &\rightarrow& z_i - 2 \\
z_{i \pm 1} &\rightarrow& z_{i \pm 1} + 1 
\end{eqnarray*}
except for the sites at the vertical wall $i = 1$ and the
open boundary $i = L$ where 
\begin{eqnarray*}
z_1 &\rightarrow& z_1 - 2 \hspace*{3.4cm} z_L \rightarrow z_L - 1 \\ 
z_2 &\rightarrow& z_2 + 1 \hspace*{3cm} z_{L-1} \rightarrow z_{L-1} + 1 .
\end{eqnarray*}
The critical slope at site $i$ is chosen randomly if 
it topples, that is $z_i^c \rightarrow 1\;\; \mbox{or}\;\; 2$.
A new metastable state is reached when $z_i \leq z_i^c$ for all $i$. 
\item Proceed to 2 and reiterate.
\end{enumerate}
Note the separation of time scales which is built into the definition
of the model. The addition of grains can only take place when the system
has settled down into a metastable state. Thus the response of the system
(the avalanches) is fast when compared with the interval between perturbations.

Metastable states $\mathcal{M}_j$ are rest states of the system.
Different metastable states occur due to the different possible
sets of critical slopes. 
Metastable states can also be divided into transient states  $\mathcal{T}_j$
and  attractor states $\mathcal{R}_j$.
Transient states are not reachable once the systems has
entered into the attractor, which are 
recurrent metastable states. The index $j$ denotes
discrete time steps associated with the long time scale of the system. 

When adding a grain to a metastable state $\mathcal{M}_j$, it evolves 
into a new metastable state $\mathcal{M}_{j+1}$ by the relaxation rules given above.  
Symbolically we write
$\mathcal{M}_j \mapsto \mathcal{M}_{j+1}$, where the arrow is a shorthand
notation for the operation of adding sand and, if necessary,
relaxing the system until it reaches a new metastable state.
Starting from, say the
empty lattice $\mathcal{T}_1$, which is a transient
state as it will never be encountered again, we have 
\begin{displaymath}
\mathcal{T}_1 \mapsto \cdots \mapsto \mathcal{T}_n \mapsto
\mathcal{R}_1 \mapsto \cdots \mapsto \mathcal{R}_{j-1} \mapsto
\mathcal{R}_{j} \mapsto \mathcal{R}_{j+1} \mapsto \cdots
\end{displaymath}
After $n$ additions and associated relaxations (if any), the system reaches the attractor of the dynamics.
For example,  
if we set $z_i^c$ independent of time, we arrive at the BTW model 
in one dimension \cite{BTW87}. 
Any grain introduced into the system after the attractor state has been 
reached will simply flow to the open boundary fall out.  
In this case, there is only one attractor state with $z_i=z_i^c\; \forall i$. 
We want to enumerate the number of different states in the attractor $\left\{\mathcal{R}_j\right\}$
for the Oslo model, where $z_i^c$ is chosen randomly between $1$ or
$2$.

\begin{figure}[h]
\includegraphics[scale = 0.75]{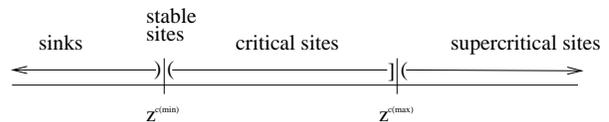}
\caption{We classify different types of sites according to the value of the local slope 
$\mathbf{z}_i$.  The above diagram shows these definitions in relation to one another,
where $z^{c(min)}$ and $z^{c(max)}$ are the minimum and maximum allowed values
for critical slopes respectively.}
\label{Fig2}
\end{figure}

We adopt the following nomenclature for sites
\begin{eqnarray}
z_i < z^{c(min)} &&\mbox{sinks,} \hspace*{0.4cm} z_i = z^{c(min)} \hspace*{0.1cm} \mbox{stable,} \nonumber \\
z^{c(min)} < z_i \leq z^{c(max)} &&\mbox{critical,} \hspace*{0.2cm}  z_i > z^{c(max)} \hspace*{0.1cm} \mbox{supercritical}. \nonumber
\end{eqnarray} 
Due to the way the problem is defined, $z_i \geq 0$ always.
Also, supercritical sites are by definition not allowed in any metastable
state.  Fig. \ref{Fig2} shows a diagrammatic relation of these
classifications.
From the relaxation rules, we find that states in the attractor 
are subject to the following two constraints. 
\begin{description}
\item [(C1)] Starting from the open boundary $i=L$, the first site ($i<L$)
that is not a stable site ($z_i\neq z^{c(min)}$) must be a critical site. 
\item [(C2)]
The first site that is not a stable state to the left of the sink must be a
critical site.  Equivalently, we cannot find a recurrent state in which two sinks
are separated only by stable sites.  The exception is at the closed boundary $i=1$, 
where the first site that is not stable must be the boundary itself.
\end{description}

\begin{figure}[h]
\includegraphics[scale = 0.65]{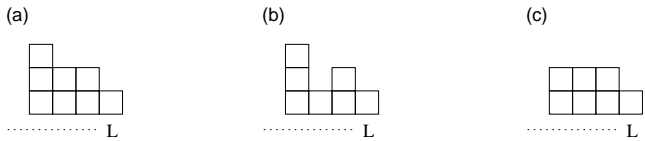}
\caption{(a) The configuration displayed violates the constraint ({\bf C1}),
as the first site that is not stable (here, $i=L-2$) next to the
open boundary $i=L$ is a sink, $z_{L-2}=0<z^{c(min)}$.
(b) and (c) are two possible preceding configurations.
These configurations do not belong to the attractor.}
\label{Fig3}
\end{figure}

\begin{figure}[h]
\includegraphics[scale = 0.3,angle=-90]{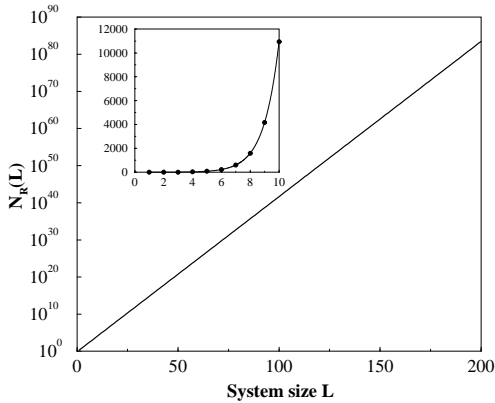}
\caption{Inset: Both simulation and analytic results are shown
for systems with $1$ to $10$
lattice sites. Results obtained by enumerating the different
metastable states in the 
attractor (filled circles) are identical to analytic results
in Eq. (\ref{result}) (solid line). 
Main figure: The analytic solution of
Eq. (\ref{result}) showing the number of states in the attractor as 
a function of system size. For a moderate system size of
$L = 195$, the number of metastable states in the
attractor exceeds $10^{81}$, the estimated number of atoms in the universe.}
\label{Fig4}
\end{figure}

We can show that these constraints must be fulfilled by assuming that
each is not true and examining how that system would enter such
as state.  For example, Fig. \ref{Fig3}(a) shows a case where
the the first rule \textbf{C1} does not hold.
The possible preceding configurations 
are displayed in Fig. \ref{Fig3}(b) and (c).  Since these
are not states in the attractor, the state displayed
in Fig. \ref{Fig3}(a) is not in 
the attractor.  One could argue likewise for the second constraint, \textbf{C2}.
Note however, that by iterating these cases, we have only shown 
that these constraints are necessary but not sufficient.  
However, we make the conjecture that these constraints are also 
sufficient for enumerating all
attractor states from exact agreement with simulations for system sizes 
$L=1$ to $10$, see Fig. \ref{Fig4}.

For a system with $L$ lattice sites, we want to find the number of states 
$N_{\mathcal{R}}(L)$ that obey the constraints above.
We can simplify the problem further by first considering 
the number states with a fixed height 
at site $i=1$, $h_1$.  We introduce the parameter 
$q=h_1-L=\sum_{i=1}^L z_i-L$ 
as it simplifies the equations involved
considerably.  Observe that the range of $q$ is $0\le q \le L$.
Let $K(L,q)$ describe the number of states in a system of $L$ sites 
and $h_1=q+L$.  $N_{\mathcal{R}}(L)$ can then be determined by
$N_{\mathcal{R}}(L)=\sum_{q=0}^L K(L,q)$.

To generate all states $\{\mathbf{z}\}$ that obey the counting constraints
for states in the attractor, consider 
the situation where we know all the allowed profiles for system sizes 
$L'<L$.  We denote the family of $n$-tuples that represent allowed profiles 
for a system of size $L$ by
$\mathcal{F}(L)$.  $N_{\mathcal{R}}(L)$ is the number
 of $n$-tuples in $\mathcal{F}(L)$.
We can generate $\mathcal{F}(L)=\{\mathbf{h}\}$
by extending $\mathcal{F}(L-1)$ by one site at the closed boundary.
This may be done using the following procedure - 
extend each $\mathbf{z}'$ to $\mathbf{z}$ by writing 
$z_{i+1}=z'_i$.
Consider one of three possible values for 
the new site next to the closed boundary, $z_1$.
$z_1=2$ for critical sites, $z_1=1$ for stable sites and $z_1=0$ for 
sinks.  
Lastly, check that the new states 
in each of the three cases obey the counting constraints ({\bf C1}) and
({\bf C2}).

Consider the three possible values for $z_1$, 
that give $\mathcal{F}(L)$ by extending $\mathcal{F}(L-1)$ corresponding to
three possible cases such that $K(L,q) = \sum_{i=1}^3K^{(i)}(L,q)$. 
\begin{enumerate}
\item
When $z_1=2$, there is a critical site next 
to the closed boundary.  
The number of allowed profiles for a system of size $L$
and parameter $q$ must be the same as the number of allowed
profiles with a system size of $L'=L-1$.  We relate
$q'$ to $q$ by observing that 
$q=h_1-L=h'_1-(L-1)+1=q'+1$.  Therefore
we assert that when $z_1=2$, $K^{(1)}(L,q)=K(L-1,q-1)$.

\item
Similarly, when $z_1=1$ corresponding to a 
stable site next to the closed boundary  
all profiles for a system size $L-1$ with $q'=q$ 
are allowed.  In this case, $K^{(2)}(L,q)=K(L-1,q)$.

\item
Finally, the situation when 
$z_1=0$ means that a sink resides next to the closed
boundary.  The counting constraints state that the 
sink next to the closed boundary must be
followed by a critical site with any number of stable sites
in between.  In this case, we consider 
$\mathbf{z}'=(0,2,\ldots)$ extending $\mathcal{F}(L-2)$, 
$\mathbf{z}'=(0,1,2,\ldots)$ extending $\mathcal{F}(L-3)$
and so on to give $L$.
This involves summing all allowed metastable
states $\mathcal{F}(L')$ with $L' \le L-2$.  
As above, we find that 
$q'=q$ for all $L'$.  Hence
\[
K^{(3)}(L,q)=\sum_{i=1}^{L-2} K(i,q).  
\]
Note that this equation is a simple summation because of
the appropriate choice of $q$.
\end{enumerate}

Summing the three cases,
\begin{eqnarray*}
K(L,q) &=&K(L-1,q-1)+K(L-1,q)+\sum_{i=1}^{L-2} K(i,q)\\
&=&\sum_{i=1}^{L-1} K(i,q)+K(L-1,q-1)\\
&=& 2 K(L-1,q)+K(L-1,q-1)-K(L-2,q-1).
\end{eqnarray*}
Hence we calculate the number of states as a function of system size by summing
through $q$ 
\begin{eqnarray*}
N_{\mathcal{R}}(L)&=&\sum_{q=0}^{L} K(L,q) \\
&=& 3 \sum_{q=0}^{L-1} K(L-1,q)-\sum_{q=0}^{L-2}K(L-2,q)\\
&=& 3N_{\mathcal{R}}(L-1)-N_{\mathcal{R}}(L-2).
\end{eqnarray*}
We have used the fact that $0 \le q \le L$.
$N_{\mathcal{R}}(L)=3N_{\mathcal{R}}(L-1)-N_{\mathcal{R}}(L-2)$ is a Fibonacci like relation. Substituting
a trial solution $N_{\mathcal{R}}(L)=c_+\lambda_+^L+c_-\lambda_-^L$, and demanding
the initial conditions $N_{\mathcal{R}}(1)=2$ and $N_{\mathcal{R}}(2)=5$, we obtain the
solution, see Fig \ref{Fig4}.
\begin{equation}
N_{\mathcal{R}}(L)=\frac{1+\sqrt{5}}{2\sqrt{5}}\left(\frac{3+\sqrt{5}}{2}\right)^L+
\frac{\sqrt{5}-1}{2\sqrt{5}}\left(\frac{3-\sqrt{5}}{2}\right)^L.
\label{result}
\end{equation} The golden
mean  $\lambda_+=\frac{3+\sqrt{5}}{2}\approx 2.618\ldots$.
Computer simulations were used to test these results for systems of size $1$ to
$10$. They were found to be in full agreement.

This result should be compared with the size of the attractor for the Abelian
Bak-Tang-Wiesenfeld sandpile determined by Dhar \cite{Dhar1}. He showed
that $N_{\mathcal{R}} = \mbox{Det} \Delta$, where $\Delta$ is the
toppling matrix. In $2D$ and for large system sizes, $N_{\mathcal{R}}$
increases exponentially fast with the number of sites in the system
$(3.21\ldots)^{L \times L}$ \cite{Creutz}.
Moreover, it was proven that the BTW model is ergodic \cite{Dhar1}.
We find, in contrast,
that the Oslo model is non-ergodic in the sense that the
states in the attractor are not visited with the same frequency.
This is most easily demonstrated by showing that the probability
of visiting stationary states in very small systems (e.g.,
$L = 2$ which has a total of $5$ states in the attractor) is
non-uniform. The number of states in the attractor
$N_{\mathcal{R}}$ does not depend on the probability distribution
with which $z_c^i$ is chosen to be $1$ or $2$, except for the trivial
cases of $P(z_c^i = 1) = 0$ or $1$, but
the distribution will affect the accessibility of a given microstate.
Finally, note that 
the Oslo model can be mapped into an interface
dragged through a random medium \cite{Maya}. Thus, the results obtained
for the Oslo model are directly applicable to the latter model as well.

{\em Possible extensions.}
A pile can be divided into an inactive zone consisting of grains that 
never will take part in the dynamics and the active zone. The width
${\lambda}_L$ of the active zone plays an important role as it is
related to various physical phenomena. E.g. the average transit times
of individual grains has been measured to increase with system size
as ${\langle T \rangle}_L \propto L {\lambda}_L \propto L^{1+\chi}$.
For the Oslo model $\chi \approx 0.25 $ while experimentally
$\chi \approx 0.5$ \cite{Oslo1,Nature}.
We believe that using a similar approach one might be able to
calculate analytically the distribution of the height of the pile at
the vertical wall $P(h_1)$ and the critical exponent $\chi$.
This would be of great importance as this exponent will be the same in
all the models belonging to the universality class of the Oslo Model.

We are grateful to S.C. Khoo,  R. Jack and {\'A}. Corral for 
their helpful comments and insights.  K. C. gratefully acknowledges
the financial support of U.K. EPSRC through Grants No. GR/R44683/01 and
GR/L95267/01.

$^*${To whom correspondence should be addressed,
E-mail: k.christensen@ic.ac.uk.}

\end{document}